\documentclass[amsmath,amssymb,aps,prb,reprint,groupedaddress]{revtex4-1}
\usepackage{graphicx}
\usepackage{dcolumn}
\usepackage{bm}
\usepackage{color}
\usepackage{multirow}

\begin{document}


\title{Anisotropic Field-Induced Gap in Quasi-One-Dimensional Antiferromagnet $\mathrm{KCuMoO_4(OH)}$}


\author{Kazuhiro Nawa$^{1,2}$}
\email{E-mail address: knawa@tagen.tohoku.ac.jp}, 
\author{Oleg Janson$^3$}
\author{Zenji Hiroi$^1$}
\affiliation{$^1$Institute for Solid State Physics, The University of Tokyo, Kashiwanoha, Kashiwa, Chiba 277-8581, Japan \\
$^2$Institute of Multidisciplinary Research for Advanced Materials, Tohoku University, Katahira, Aoba-ku, Sendai 980-8577, Japan \\
$^3$Institut f\"{u}r Festk\"{o}rperphysik, TU Wien, Wiedner Hauptstra\ss e 8-10, 1040 Vienna, Austria}

\date{\today}

\begin{abstract}
We investigated magnetic and thermodynamic properties of $S$ = 1/2 quasi-one-dimensional antiferromagnet $\mathrm{KCuMoO_4(OH)}$ through single crystalline magnetization and heat capacity measurements.
At zero field, it behaves as a uniform $S$ = 1/2 Heisenberg antiferromagnet with $J$ = 238~K,
and exhibitsa canted antiferromagnetism below $T_N$ = 1.52~K.
In addition, a magnetic field $H$ induces the anisotropy in magnetization and opens a gap in the spin excitation spectrum.
These properties are understood in terms of an effective staggered field induced by staggered $g$-tensors and Dzyaloshinsky-Moriya (DM) interactions.
Temperature-dependencies of the heat capacity and their field variations are consistent with those expected for quantum sine-Gordon model, indicating that spin excitations consist of soliton, anti-soliton and breather modes.
From field-dependencies of the soliton mass, the staggered field normalized by the uniform field $c_\mathrm{s}$ is estimated as 0.041, 0.174, and 0.030, for $H \parallel a$, $b$, and $c$, respectively.
Such a large variation of $c_\mathrm{s}$ is understood as the combination of staggered $g$-tensors and DM interactions which induce the staggered field in the opposite direction for $H \parallel a$ and $c$ but almost the same direction for $H \parallel b$ at each Cu site.
\end{abstract}


\maketitle

\section{\label{introduction}Introduction} 
A one-dimensional magnet is a paradigmatic model which exhibits an intriguing magnetism
in spite of its model simplicity\cite{SL}.
After the Ising model was solved analytically\cite{Lenz, Ising} and 
a method to find the exact quantum mechanical ground state of a Heisenberg antiferromagnet was developed\cite{Bethe}, 
extensive studies have revealed unconventional ground states and quantum critical phenomena in various one-dimensional magnets\cite{SL, Haldene, IsingAn, alternatechain, critical}.
In particular, a one-dimensional Heisenberg antiferromagnet with $S$ = 1/2 does not exhibit
a magnetic order at the ground state
but has continuous spin excitations which are gapless at wave vector $q$ = 0 and $\pi$\cite{dCP}.
They can be regarded as two different domain-wall motions, named spinons\cite{spinon},
and thus this spin excitation spectrum is called two spinon continuum.
In fact, such excitation was observed in one-dimensional antiferromagnet $\mathrm{CuCl_2 \cdot 2N(C_5D_5)}$\cite{1Dinelastic} and $\mathrm{KCuF_3}$\cite{1Dinelastic2}.

Furthermore, the effect of a magnetic field on spinon excitations was investigated through inelastic neutron scattering experiments on Cu-benzoate\cite{Cubenzoate1, Cubenzoate2, Cubenzoate3}.
Although theories indicated the appearance of additional \textit{gapless} excitations at 
$q$= $\pm 2 \pi M/(g \mu_\mathrm{B})$ and $\pi \pm 2 \pi M/(g \mu_\mathrm{B})$
($M$ is magnetization per one spin)\cite{1DAF, 1DAF2},
\textit{gapped} excitations were detected at the corresponding $q$-vector\cite{Cubenzoate3}.
This discrepancy is well understood by staggered $g$-tensors and Dzyaloshinsky–Moriya (DM) interactions,
which are also effective in Cu-benzoate\cite{Cubenzoate3, 1DstagDMchain, 1DstagDMchain2}.
In the presence of an external magnetic field, they induce a staggered field, which breaks rotation symmetry perpendicular to the magnetic field.

\begin{figure}[t]
\centering
\includegraphics[width=8cm]{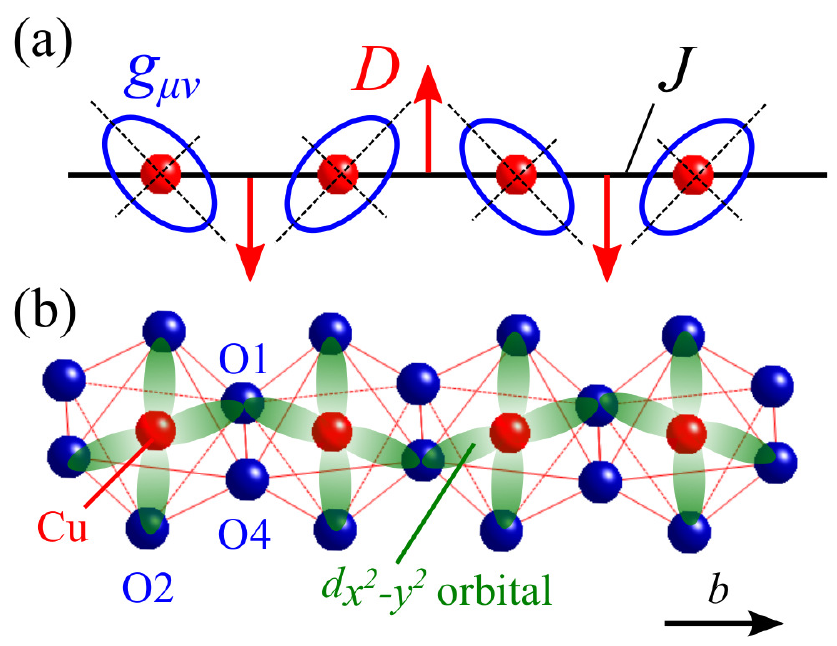}
\caption{\label{fig:struct}
(a) Schematic view of a one-dimensional antiferromagnet with staggered $g$-tensors (blue circles) and DM interactions (red arrows). A dashed line represents principal axes for each $g$-tensor.
Directions of its principal axes and DM vectors are chosen so that they match those of $\mathrm{KCuMoO_4(OH)}$.
(b) Cu and O atoms of $\mathrm{CuO_6}$ octahedra in $\mathrm{KCuMoO_4(OH)}$.
Green ellipses represent $x^2-y^2$ orbitals carrying $S$ = 1/2 spins.
}
\end{figure}

\begin{figure}[t]
\centering
\includegraphics[width=8cm]{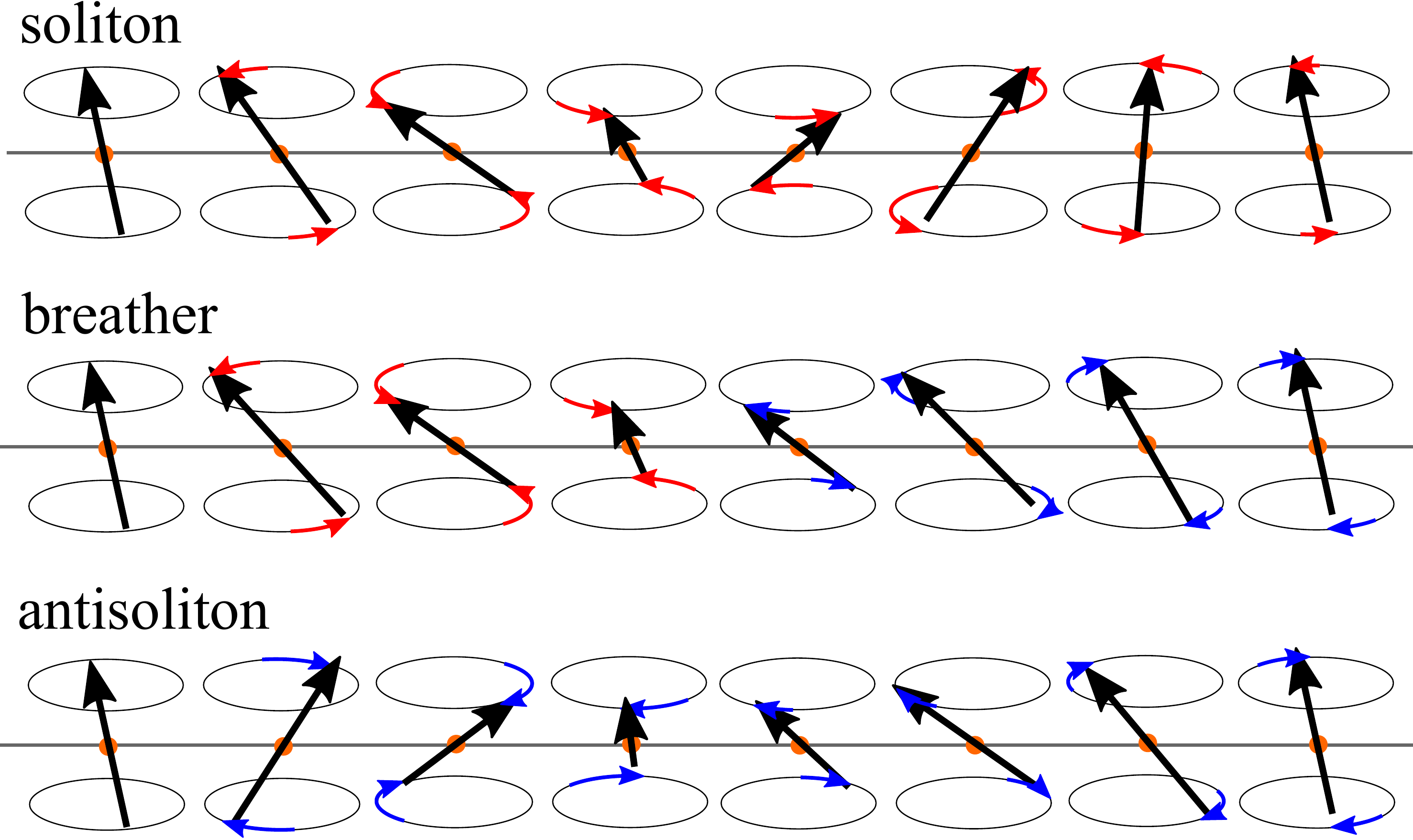}
\caption{\label{breather} Schematic view of soliton, breather, antisoliton excitations expected in a one-dimensional antiferromegnetic chain with staggered $g$-tensors and DM interactions.
Red (blue) arrows indicate that transverse spin components are twisted counterclockwise (clockwise).}
\end{figure}

To explain this in detail, we start from a model Hamiltonian: 
\begin{equation}
\begin{split}
H &= \sum_i \{ J \mathbf{S}_i \cdot \mathbf{S}_{i+1} + (-1)^{i+1} \mathbf{D}_0 \cdot (\mathbf{S}_{i} \times \mathbf{S}_{i+1}) \\
&\ \ \ - \mu_\mathrm{B} \mathbf{H} (\mathbf{g}_\mathrm{u} + (-1)^i \mathbf{g}_\mathrm{s}) \mathbf{S}_i \}, \label{basicHamiltonian}
\end{split}
\end{equation}
where $\mu_\mathrm{B}$ represents Bohr magneton.
The $\mathbf{g}_\mathrm{u}$ and $\mathbf{g}_\mathrm{s}$ are uniform and staggered components of $g$-tensor defined on each Cu atom, respectively.
In addition, $\mathbf{D}_0$ is a vector which represents a DM interaction defined on each bond connecting two Cu atoms.
The schematic view of staggered $g$-tensors and alternating DM vectors is shown in Fig.~\ref{fig:struct}(a).
Introducing a spin rotation around the direction of DM vectors modifies Hamiltonian \eqref{basicHamiltonian} into an effective model as\cite{sign}:
\begin{equation}
\begin{split}
H &= \sum_i \{ J \mathbf{S}_i \cdot \mathbf{S}_{i+1} - \mathbf{h}_\mathrm{u} \cdot \mathbf{S}_i - (-1)^i \mathbf{h}_\mathrm{s} \cdot \mathbf{S}_i \}, \\
\mathbf{h}_\mathrm{u} &\sim \mu_\mathrm{B} \mathbf{g}_\mathrm{u} \mathbf{H}, \ \mathbf{h}_\mathrm{s} \sim \mu_\mathrm{B} \Big( \mathbf{g}_\mathrm{s} \mathbf{H} - \frac{\mathbf{D}_0}{2J} \times \mathbf{g}_\mathrm{u} \mathbf{H} \Big), \label{stagfield}
\end{split}
\end{equation}
where $\mathbf{h}_\mathrm{u}$ and $\mathbf{h}_\mathrm{s}$ represent a uniform field and a staggered field, respectively.
Since $\mathbf{h}_\mathrm{u}$ and $\mathbf{h}_\mathrm{s}$ are almost perpendicular to each other,
we consider a simple case when $\mathbf{h}_\mathrm{u}$ and $\mathbf{h}_\mathrm{s}$ are aligned along the $z$ and $x$-axis, respectively, as
\begin{equation}
H = \sum_i \{ J \mathbf{S}_i \cdot \mathbf{S}_{i+1} - h_\mathrm{u} S_i^{z} - (-1)^i h_\mathrm{s} S_i^{x} \}. \label{stagfield2}
\end{equation}
The low-energy behavior of Hamiltonian \eqref{stagfield2} is well described by sine-Gordon model with a Lagrangian density $L = (\partial_\mu \phi)^2/2 + C h \cos(2\pi R \tilde{\phi})$,
where $\phi$, $\tilde{\phi}$, $R$, $h$, and $C$ represent a boson field, a dual field, compactification radius, Planck constant and an arbitrary number\cite{1DstagDMchain, 1DstagDMchain2}. 
This model leads to the low-energy excitations of soliton, anti-soliton, and breather modes,
which are schematically illustrated in Fig.~\ref{breather}.zero twist.
This mechanism has been shown to underlie excitations in several one-dimensional antiferromagnets such as
Cu-benzoate\cite{Cubenzoate1, Cubenzoate2, Cubenzoate3, Cubenzoate4, Cubenzoate5, Cubenzoate6},
$\mathrm{CuCl_2\cdot 2(CH_3)_2SO}$\cite{CDC, CDC2, CDC3},
$\mathrm{Yb_4As_3}$\cite{Yb4As3, Yb4As3_1, Yb4As3_2, Yb4As3_3},
$[\mathrm{PM \cdot Cu(NO_3)_2 \cdot (H_2O)_2]_n}$] (PM = pyrimidine)\cite{PMCH, PMCH2},
$\mathrm{KCuGaF_6}$\cite{KCuGaF6, KCuGaF6_2, KCuGaF6_3, KCuGaF6_4},
and $\mathrm{CuSe_2O_5}$\cite{CuSe2O5, CuSe2O5_2}
through magnetization, heat capacity, inelastic neutron scattering, and electron spin resonance measurements,
as summarized in Table~\ref{compounds}. 

\begin{table}[t]
\caption{
\label{compounds} Model compounds of a quasi-one-dimensional antiferromagnet with staggered $g$-tensors and DM interactions.
Crystal symmetry, the nearest-neighbor exchange $J$, a transition temperature $T_\mathrm{N}$, and
a staggered field normalized by a uniform field $c_\mathrm{s} \equiv |h_\mathrm{s}/h_\mathrm{u}|$ are listed. 
$\mathrm{CuCl_2\cdot 2(CH_3)_2SO}$ and $\mathrm{[PM\cdot Cu(NO_3)_2\cdot (H_2O)_2]_n}.$ (PM = pyrimidine) are abbreviated as CDC and Cu-PM, respectively.
The $a''$ and $c''$ represent crystal axes individually defined in the references.}
\label{coeff}
\begin{ruledtabular}
\begin{tabular}{cccc}
Compound (symmetry) & $J$ (K) & $T_N$ (K) & $c_\mathrm{s}$ \\ \hline
CDC\cite{CDC, CDC2, CDC3} ($Pnma$) & 17 & 0.93 & $H \parallel a$ : 0.045 \\
 & & & $H \parallel b$ : 0 \\
 & & & $H \parallel c$ : 0.090 \\
Cu-benzoate\cite{Cubenzoate1, Cubenzoate2, Cubenzoate3, HC, HC_2, Cubenzoate4, Cubenzoate5, Cubenzoate6}  & 18 & $<$0.02 & $H \parallel a''$ : $\sim$0 \\
($I2/c$) & & & $H \parallel c''$ : 0.111 \\
$\mathrm{Yb_4As_3}$\cite{Yb4As3, Yb4As3_1, Yb4As3_2, Yb4As3_3} ($R3c$) & 26 & $<$0.045 & (poly) 0.27  \\
Cu-PM\cite{PMCH, PMCH2} ($C2/c$) & 36 & $<$0.38 & $H \parallel a''$ : $\sim$0 \\
  & & & $H \parallel b$ : 0.07--0.09 \\
  & & & $H \parallel c''$ : 0.10--0.12 \\
$\mathrm{KCuGaF_6}$\cite{KCuGaF6, KCuGaF6_2, KCuGaF6_3, KCuGaF6_4} ($P2_1/c$) & 103 & $<$0.05 & $H \parallel a$ : 0.031  \\
 & & & $H \parallel b$ : 0.160 \\
 & & & $H \parallel c$ : 0.178 \\
$\mathrm{CuSe_2O_5}$\cite{CuSe2O5, CuSe2O5_2} ($C2/c$) & 157 & 17 & $H \parallel c$ : 0.08  \\
$\mathrm{KCuMoO_4(OH)}$ ($Pnma$) & 238 & 1.52 & $H \parallel a$ : 0.041 \\
 & & & $H \parallel b$ : 0.174 \\
 & & & $H \parallel c$ : 0.030 \\
\end{tabular}
\end{ruledtabular}
\end{table}

$\mathrm{KCuMoO_4(OH)}$ is a new $S$ = 1/2 quasi-one-dimensional antiferromagnet with
staggered DM interactions\cite{KCuMoO4OH}.
Its local structure around Cu atoms is shown in Fig.~\ref{fig:struct}(b).
It crystallizes in an orthorhombic structure with the space group $Pnma$ and
consists of edge-sharing $\mathrm{CuO_6}$ octahedra, forming a chain of $S = 1/2$ along the $b$-axis.
The key element of its magnetism is a staggered arrangement of $d_{x^2-y^2}$ orbitals occupied by unpaired spins, which are illustrated as green ellipses linked along the chain in Fig.~\ref{fig:struct}(b).
Due to superexchange couplings through Cu--O1--Cu paths,
the nearest-neighbor exchange $J$ becomes largely antiferromagnetic, $J$ = 238 K.
Note that a different arrangement of $d_{x^2-y^2}$ orbitals can make $J$ ferromagnetic,
such as that in $\mathrm{NaCuMoO_4(OH)}$\cite{NaCuMoO4OH, NaCuMoO4OH2} where $d_{x^2-y^2}$ orbitals are linked by two bridging oxygen atoms.
Staggered $g$-tensors and DM interactions in $\mathrm{KCuMoO_4(OH)}$ give rise to
enhanced magnetic susceptibility at low temperatures,
large anisotropy in magnetization,
and an exponential temperature-dependence in heat capacity,
indicating the presence of the field-induced gapped excitations.
In fact, these behaviors have been also observed in the other model compounds\cite{Cubenzoate2, Cubenzoate3, Yb4As3_2, PMCH, KCuGaF6, KCuGaF6_3}.

$\mathrm{KCuMoO_4(OH)}$ has a few advantages for investigating the nature of the low-energy excitations.
First, a soliton mass can be determined more easily since the temperature range necessary to investigate the excitations becomes higher owing to large $J$,
compared with most of previous model compounds with small $J$,
which are suitable for investigating the effect of magnetic fields.
Secondly, its higher crystal symmetry makes its magnetic character simpler
than that in other compounds such as Cu-benzoate ($I2/c$)\cite{Cubenzoate1}.
Due to the crystal symmetry of $Pnma$, a single Cu chain includes two different Cu sites related to each other by
a mirror symmetry that bisects a Cu--Cu bond and a twofold screw axis that passes through the Cu atoms.
They require 
the DM vector to align perpendicular to the chain and to alternate along the chain direction, respectively\cite{KCuMoO4OH}.
Thus, extra interactions such as uniform components of the DM vector are absent and
the magnetic interactions are very well described by Hamiltonian \eqref{basicHamiltonian}.
In this study, we take advantage of these characteristics and discuss the field-dependence of soliton formation through detailed magnetization and heat capacity measurements. 

The paper is organized as follows.
Experimental results on magnetization and heat capacity measurements are described in Sec.~\ref{measurements}.
The magnetic and thermodynamic properties are consistent with those expected for sine-Gordon model,
indicating that a magnetic field induces the soliton, antisoliton, and breather modes in the spin-excitation spectrum.
The magnitude of the staggered field normalized by the uniform field is estimated from field-dependencies of the soliton mass.
In Sec.~\ref{DFT}, we present results of DFT calculations, which are consistent with the experimental results.
In Sec.~\ref{estimation}, we discuss the origin of the staggered field based on symmetry properties of $g$-tensors and DM vectors.
The reason why $c_\mathrm{s}$ varies largely with a field-direction is that staggered $g$-tensors and Dzyaloshinsky–Moriya interactions induce $\mathbf{h}_\mathrm{s}$
in the opposite direction for $H \parallel a$ and $c$ but almost the same direction for $H \parallel b$.
Thus, net $h_\mathrm{s}$ becomes small for $H \parallel a$ and $c$ but large for $H \parallel b$.
A summary is presented in Sec.~\ref{conclusions}.

\section{\label{measurements}Magnetization and heat capacity measurements}
Single crystalline samples with a typical size of 0.5 $\times$ 0.5 $\times$ 0.5 mm$^3$
were prepared by hydrothermal method\cite{KCuMoO4OH}. 
Its magnetization and heat capacity were measured in a SQUID magnetometer (Quantum Design MPMS)
and by the relaxation method (Quantum Design PPMS), respectively.
A $^3$He insert was used for the measurements below 2 K.
We carefully co-aligned 8 pieces of single crystalline samples to investigate the variation of the magnetization and heat capacity depending on a magnetic field direction.
However, an aggregate of randomly oriented single crystals is used for a magnetization measurement below 2 K.

\begin{figure}[t]
\centering
\includegraphics[width=8cm]{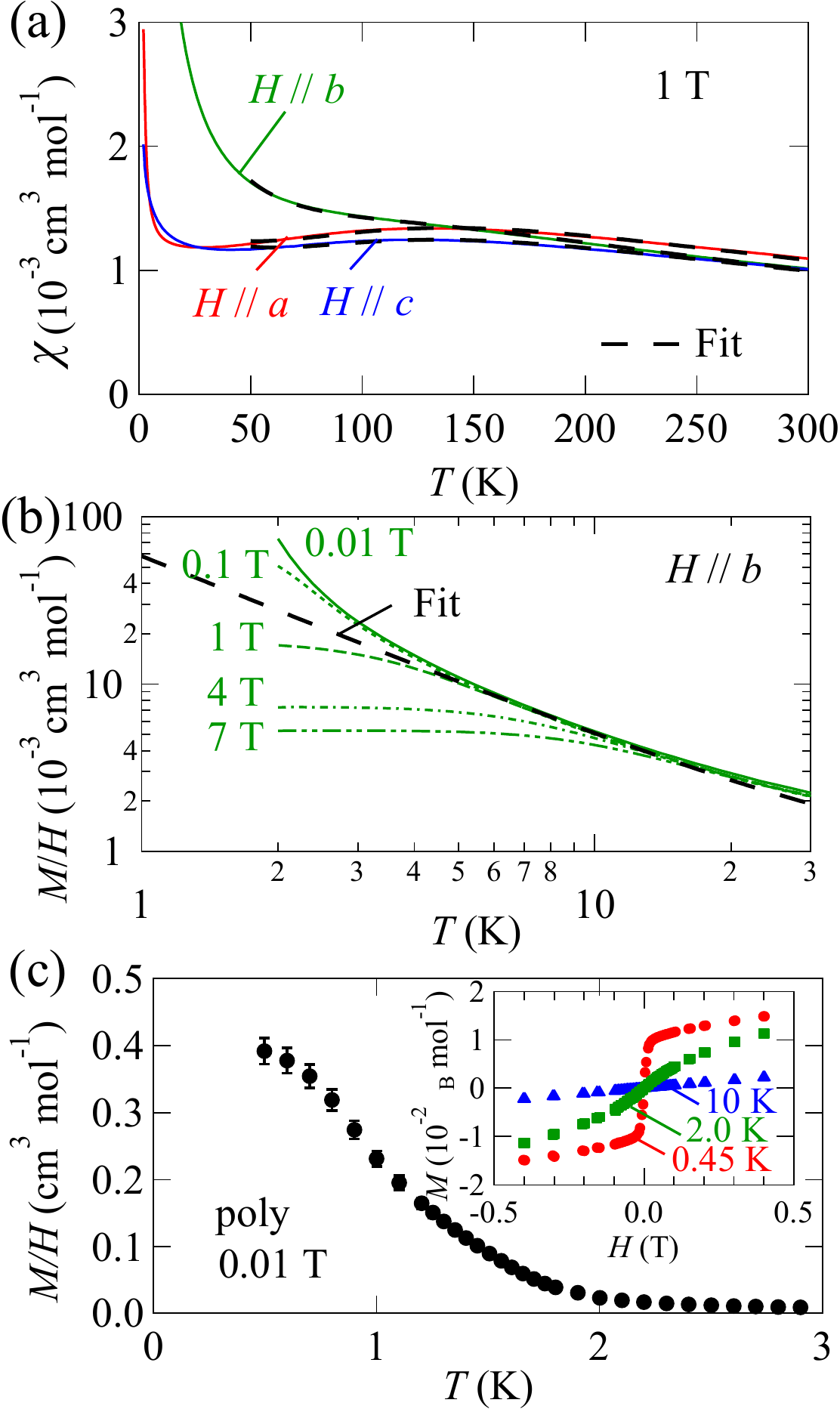}
\caption{\label{fig:chi} (a) Temperature-dependencies of the magnetic susceptibility above 2~K for $H \parallel a$, $b$, and $c$ in a field of 1 T.
Dashed curves represent a fit to the Eqs.~\eqref{unichi}.
(b) Double-logarithmic plot of the magnetic susceptibility between 2--30~K. The dashed curve represents a fit to the Eqs.~\eqref{stagchi}.
(c) The magnetic susceptibility below 2~K and magnetization curves (shown in the inset) of randomly oriented single crystals.}
\end{figure}

Temperature-dependencies of magnetic susceptibility $\chi_i$ for $H \parallel a$, $b$, and $c$, ($i$ = $a$, $b$, and $c$, respectively) are shown in Fig.~\ref{fig:chi}(a).
A broad hump appears around 150 K in $\chi_a$ and $\chi_c$, and a broad shoulder is present in $\chi_b$,
indicating a one-dimensional character.
At low temperatures, $\chi$ largely increases with strong field-direction dependence as shown in Fig.~\ref{fig:chi}(b).
This feature is characteristic for a one-dimensional chain with staggered g-tensors and DM interactions,
and usually not caused by defects or magnetic impurities\cite{KCuMoO4OH}.
To estimate the magnitude of the antiferromagnetic exchange,
$\chi_i$ in the temperature range of 50--300~K is fitted to the sum of a one-dimensional antiferromagnetic chain model\cite{1D} and a Curie contribution as
\begin{equation}
\chi_\mathrm{fit} = \chi_\mathrm{u} + \frac{C_i}{T} + \chi_0, \ \ \
\chi_\mathrm{u} = \frac{N g_{ii}^2 \mu_\mathrm{B}^2}{4 k_B T} F\Bigg( \frac{J}{k_B T} \Bigg), \label{unichi}
\end{equation}
where $F(x)$ 
is a [5, 6]-Pad\'e-approximant-based parametrized solution for the spin-1/2 Heisenberg chain valid in the experimentally studied temperature range\cite{1D}.
The Curie contribution is introduced to represent a large increase of $\chi_i$ due to the staggered susceptibility\cite{PMCH, KCuGaF6, KCuGaF6_2}, which is enough for the rough estimation of $J$.
Totally, seven fitting parameters are included:
$J$ independent on the field direction, and $g_{ii}$, $C_i$ varying with the field direction ($i$ = $a$, $b$, or $c$).
In the fit, we keep a temperature-independent contribution $\chi_0$ to $-9.1 \times10^{-5}$ cm$^3$ mol$^{-1}$, which is estimated from Pascal constants of the constituent atoms.
If $\chi_0$ is added as an adjustable parameter, the fit yields $\chi_0$ = 1.6 $\times10^{-4}$ cm$^3$ mol$^{-1}$, which is too large for Van-Vleck paramagnetism for Cu$^{2+}$;
$\chi_0$ tends to be overestimated because of the weak temperature-dependence of $\chi$ between 50 and 300~K.
Fitting curves well reproduce the experimental data, as shown as dashed curves in Fig.~\ref{fig:chi}(a).
The fit yields $J$ = 232(8) K, $g_{aa}$ = 2.40(3), $g_{bb}$ = 2.18(3), and $g_{cc}$ = 2.29(3).
The parameters obtained from the fit are listed in Table~\ref{tab:param}.

\begin{table}[t]
\caption{\label{tab:param} Nearest-neighbor exchange $J$, diagonal elements of $g$-tensor $g_{ii}$,
and a Curie term $C_i$ ($i$~= $a$, $b$, and $c$) estimated using a fit to the magnetic susceptibility. 
The isotropic exchange $J$ is independent on the field direction. A temperature independent susceptibility $\chi_0$ of $-9.1 \times10^{-5}$ cm$^3$ mol$^{-1}$ is also used in the fit.}
\begin{ruledtabular}
\begin{tabular}{cccc}
Field direction & $J$ (K) & $g_{ii}$ & $C_i$ ($10^{-2}$ cm$^3$~K$^{-1}$ mol$^{-1}$)
\\ \hline
$H \parallel a$ &  & 2.40(3) & 1.0(4) \\
$H \parallel b$ & 232(8) & 2.18(3) & 4.4(3) \\
$H \parallel c$ &  & 2.29(3) & 1.2(5) \\
\end{tabular}
\end{ruledtabular}
\end{table}

The magnetization also indicates the occurrence of canted-antiferromagnetism in a low temperature.
Figure~\ref{fig:chi}(c) shows temperature-dependence of magnetization $M/H$ of randomly oriented single crystals.
$M/H$ largely increases below 1.7~K, and tends to saturate at 0.5~K.
Thus, we can deduce that a magnetic transition to a canted-antiferromagnetic state is present around 1.7~K.
This is consistent with magnetization curves shown in the inset of Fig.~\ref{fig:chi}(c).
Spontaneous magnetization of 0.01 $\mu_\mathrm{B}$/Cu is present in 0.45~K while it is absent at 10 and 2~K.
Generally, DM interactions induce effective magnetic fields perpendicular to the DM vector.
Thus, staggered DM interactions induce staggered effective magnetic fields, leading to two-sublattice canted antiferromagnetism.
A large increase in $\chi$ at low temperatures results from a short-range order.

\begin{figure}[t]
\centering
\includegraphics[width=8cm]{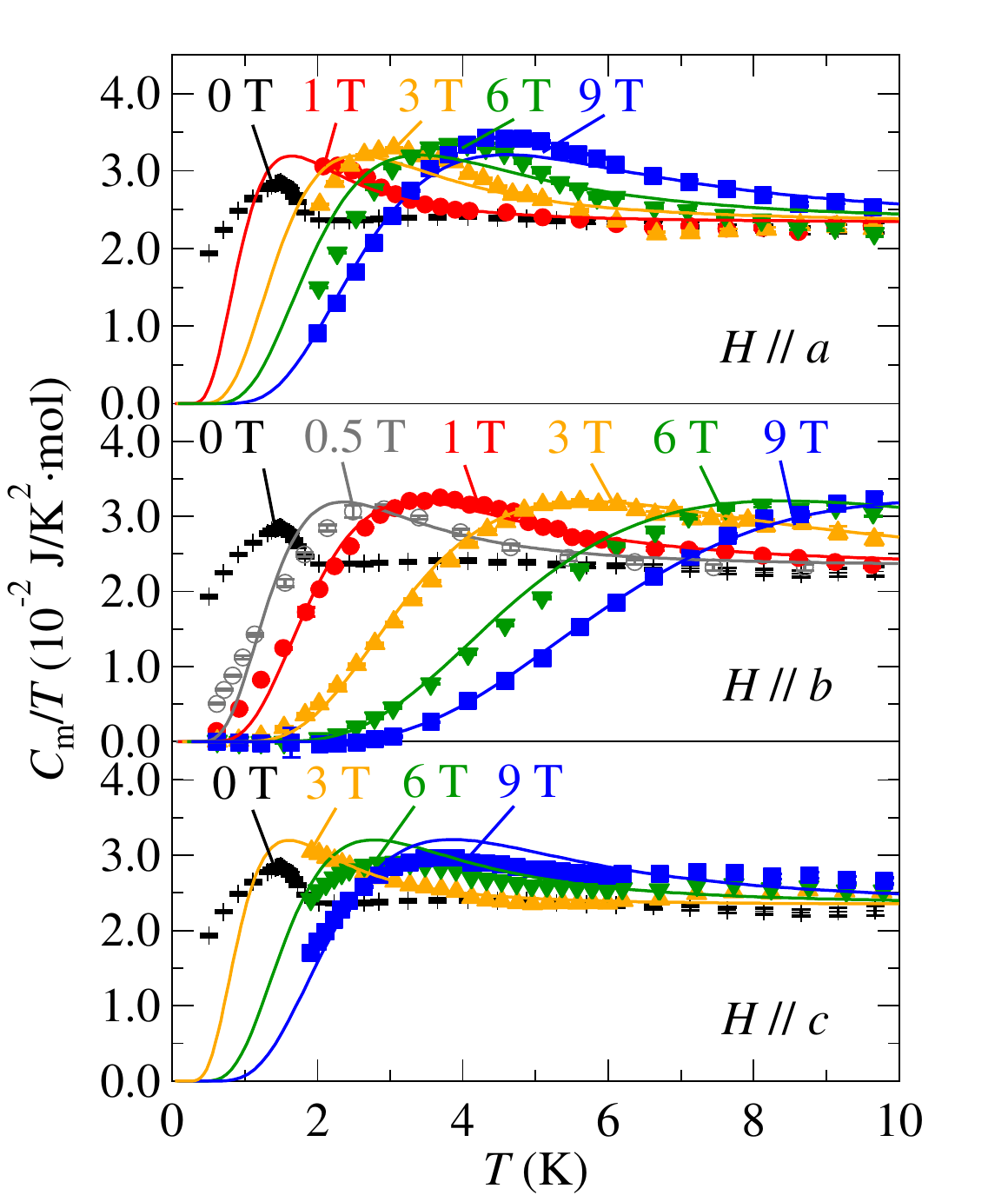}
\caption{\label{fig:HC} Temperature-dependencies of the magnetic heat capacity divided by temperature. Magnetic fields of 0--9 T are applied along $H \parallel a$, $b$, and $c$ axes. Solid curves represent a fit to the calculated curves from Bethe ansatz\cite{KCuGaF6_2}.}
\end{figure}

Heat capacity measurements also support the one-dimensional character indicated by the magnetization measurements.
Temperature-dependencies of magnetic heat capacity divided by temperature $C_m/T$ is shown in Fig.~\ref{fig:HC}.
The $C_m$ is derived as follows: first, the total heat capacity at zero field is fitted by the sum of magnetic contributions given by a one-dimensional antiferromagnetic model\cite{1D}
and phonon contributions given by the sum of Debye and Einstein model.
Then the phonon contributions are subtracted from the total heat capacity.
In zero field, $C_m/T$ becomes almost constant between 2--10~K.
This behavior is consistent with the temperature-dependence of a one-dimensional antiferromagnet in a low temperature limit, given by $C_m/T = 2R/3J$\cite{1D}.
The $J$ is estimated as $J = 238(1)$~K by using this relation\cite{KCuMoO4OH}, which agrees well with that estimated from $\chi$.
In addition, $T_\mathrm{N}$ is determined as 1.52 K from a small peak in $C_m/T$, which is also consistent with that expected from $M/H$.
From the $T_\mathrm{N}/J$ ratio of $6 \times 10^{-3}$, the magnitude of an interchain exchange $J'$ can be roughly estimated as $J' \sim 2.6 \times 10^{-3} J \sim 0.6$~K\cite{TN}.
The ratio $J'/J$ is several times larger than that of Cu-benzoate\cite{Cubenzoate5} ($J'/J < 4 \times 10^{-4}$)
and $\mathrm{Sr_2CuO_3}$\cite{Sr2CuO3, Sr2CuO3_2} ($J'/J \sim 7 \times 10^{-4}$),
and almost the same as that of $\mathrm{Ca_2CuO_3}$\cite{Sr2CuO3_2, Ca2CuO3} ($J'/J \sim 2 \times 10^{-3}$),
indicating a good one-dimensionality of $\mathrm{KCuMoO_4(OH)}$.

Next we discuss magnetization and heat capacity measurements in external magnetic fields.
According to the sine-Gordon theory, they induce staggered magnetic fields
and thus lead to a large increase in $\chi$ at low temperatures. 
The $\chi$ consists of a uniform susceptibility $\chi_\mathrm{u}$ given in Eqs.~\eqref{unichi} and a staggered susceptibility $\chi_\mathrm{s}$.
For $T \ll J$, $\chi$ is described as\cite{1DstagDMchain2}:
\begin{equation}
\chi \sim \chi_\mathrm{u} + c_\mathrm{s}^2 \chi_\mathrm{s}, \ \ \
\chi_\mathrm{s} \sim 0.278 \ \frac{N g^2 \mu_\mathrm{B}^2}{k_B T} \sqrt{ \ln \Bigg( \frac{J}{k_B T} \Bigg) }, \label{stagchi}
\end{equation}
where $c_\mathrm{s}$, $N$ and $k_\mathrm{B}$ are a ratio of a staggered field to a uniform field $|h_\mathrm{s}/h_\mathrm{u}|$, Avogadro number, and Boltzmann constant, respectively.
While $\chi_\mathrm{u}$ becomes almost constant, $\chi_\mathrm{s}$ increases approximately proportionally to $1/T$.
Figure~\ref{fig:chi}(b) is a double-logarithmic plot of $M/H$ versus $T$.
As expected from Eqs.~\eqref{stagchi}, the $M/H$--$T$ plot measured in low fields behaves as a straight line 
in the temperature range where $T_N \ll T \ll J$ and $\chi_\mathrm{u} \ll \chi_\mathrm{s}$ hold.
The $M/H$ between 5 and 15 K in a magnetic field of $H = 0.01$~T is well fitted to Eqs.~\eqref{stagchi},
as shown by a dashed curve in Fig.~\ref{fig:chi}(b).
By fixing $J$ and $g_{bb}$ to those obtained from the fit of $\chi$ in a high temperature (see Table~\ref{tab:param}), the fit yields $c_\mathrm{s} = 0.227(5)$.
In much higher fields, $M/H$ is decreased and becomes constant at low temperatures since the gap cuts off thermal excitations\cite{1DstagDMchain2}.

The antiferromagnetic transition is suppressed rapidly by a small magnetic field, as revealed from heat capacity measurements.
As shown in Fig.~\ref{fig:HC}, the peak present in zero field disappears and instead an exponential temperature-dependence becomes apparent as a magnetic field increases.
These behaviors can be understood as an analogy with those in a simple ferromagnetic system: 
a finite \textit{uniform} magnetic field smears out a ferromagnetic transition and induces a gap in a magnetic excitation spectrum, since spontaneous symmetry breaking is lost in a nonzero \textit{uniform} field.
A crossover from a paramagnetic phase to a forced-ferromagnetic phase occurs as a temperature is decreased.
Similarly, in a weakly coupled antiferromagnetic chain system, a finite \textit{staggered} magnetic field can smear out the transition to a two-sublattice antiferromagnetic order and opens a gap,
because of the absence of spontaneous symmetry breaking in a nonzero \textit{staggered} field.
Thus, a crossover from a paramagnetic phase to a forced-antiferromagnetic phase occurs in a finite \textit{staggered} field which is induced by a finite external field (Eq.~\eqref{stagfield}).
More precisely, the antiferromagnetic order at zero field can persist in a very small staggered field
if interchain interactions produce effective fields which compete with the staggered field\cite{CMFT}.
This is because symmetry of the antiferromagnetic order differs from that of the forced-antiferromagnetic state.
An external field necessary to suppress the antiferromagnetic order is estimated as an order of 0.01 T from $J$ = 238 K and $J'/J = 2.6 \times 10^{-3}$, according to the formula based on a chain mean field theory\cite{CMFT}.

The presence of the field-induced soliton, antisoliton, and breather excitations is evidenced by an exponential temperature-dependence of the heat capacity.
Their magnitude can be determined from the temperature-dependence of $C_m$.
By applying an iterative procedure to nonlinear integral equation of sine-Gordon free energy,
$C_m$ is approximately given by an analytical form which holds at a low temperature as\cite{HC},
\begin{equation}
\begin{split}
C_m &\sim 2C_\mathrm{s} + \sum_{\alpha = 1}^{1/\xi} C_\alpha, \\
C_\mathrm{s} &= \frac{M_0 R}{\sqrt{2 \pi} J v} \Bigg\{ 1 + \frac{k_B T}{M_0} + \Bigg( \frac{k_B T}{M_0} \Bigg)^2 \Bigg\}  \Bigg( \frac{M_0}{k_B T} \Bigg)^\frac{3}{2} \\
&\ \ \ \ \ \ \times \exp \Bigg( -\frac{M_0}{k_B T} \Bigg), \\
C_\alpha &\equiv \frac{M_\alpha R}{\sqrt{2 \pi} J v} \Bigg\{ 1 + \frac{k_B T}{M_\alpha} + \Bigg( \frac{k_B T}{M_\alpha} \Bigg)^2 \Bigg\} \Bigg( \frac{M_\alpha}{k_B T} \Bigg)^\frac{3}{2} \\
&\ \ \ \times \exp \Bigg( -\frac{M_\alpha}{k_B T} \Bigg). \label{C_SGmodel}
\end{split}
\end{equation}
A velocity of soliton $v$ and a parameter $\xi$ defined from compactification radius
are determined numerically from Bethe ansatz integral equations\cite{1DstagDMchain2, HC_2}.
In the present case, where the magnetic field is much smaller than $J$,
the parameters are $v \sim \pi/2$ and $1/\xi = 3 \sim 3.3$.
The $C_m$ includes contributions from five different low-energy excitations:
soliton ($M_0$ in Eqs.~\eqref{C_SGmodel}), antisoliton ($M_0$), and 3 breathers ($M_1, M_2, M_3$),
which are related to each other by the equation,
\begin{equation}
M_\alpha = 2 M_0 \sin\Big( \frac{n \pi \xi}{2} \Big) \ (\alpha = 1, 2, \cdots, [1/\xi]), \label{relation_M}
\end{equation}
where [...] denotes a floor function.
Thus, $M_0$ is the only free parameter included in Eqs.~\eqref{C_SGmodel}.

To estimate $M_0$ more accurately, we fit $C_m$ by a function which is calculated
by the thermodynamic Bethe ansatz method\cite{KCuGaF6_2}.
The fitting curves are shown as solid curves in Fig.~\ref{fig:HC}.
While fits in low magnetic fields are not so good, those in high magnetic fields agree well with the experimental data.
This is because the fit is based on a purely one-dimensional system without interchain interactions,
while in fact they are present in a real material.
Magnon dispersions of 0.6 K due to interchain interactions can be critical in a low magnetic field,
when the gap energy is below 5 K.
However, they become almost negligible in a high magnetic field when the gap energy exceeds 10 K.
The soliton mass $M_0$ at 9 T is estimated as 11(2), 27(1), and 9.3(15) K for $H \parallel a$, $b$, and $c$, respectively.
The $M_0$ of 27 K is much larger than 5.5 K for Cu-benzoate ($H \parallel c''$)\cite{Cubenzoate3, HC},
and even a little larger than 24--25 K for $\mathrm{KCuGaF_6}$ ($H \parallel c$)\cite{KCuGaF6_4}.

\begin{figure}[t]
\centering
\includegraphics[width=8cm]{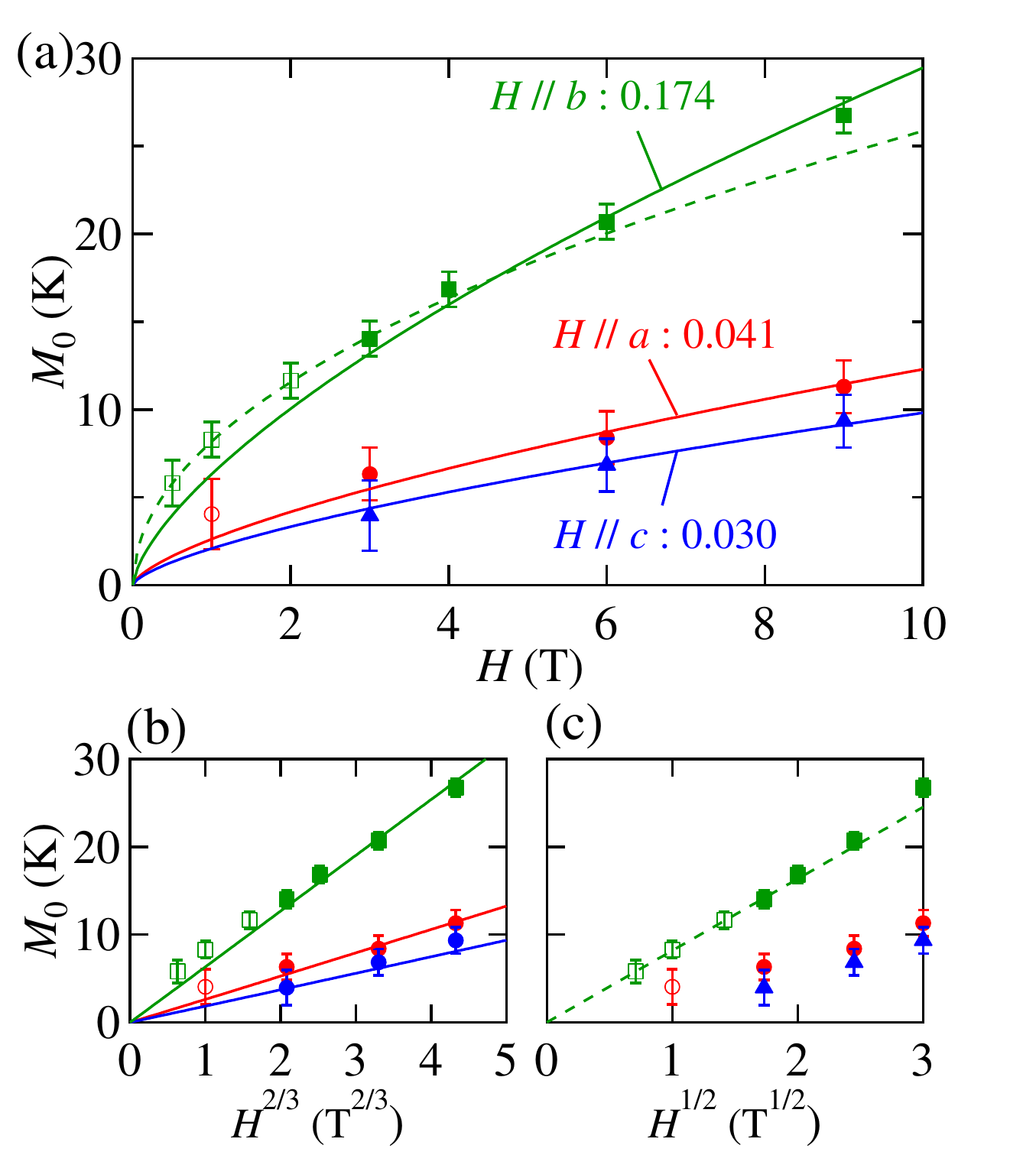}
\caption{\label{fig:Gap} (a) Field-dependencies of the soliton mass $M_0$ for $H \parallel a$, $b$, and $c$.
The solid curve represents a fit to Eq.~\eqref{M0}, and $c_\mathrm{s}$ estimated from the fit is shown close to the fitting curve.
A dashed curve is a fit to Eq.~\eqref{M00}, which is derived from a mean field theory.
The former and the latter fits are performed in a field region of 0--3 T and 3--9 T (filled symbols), respectively.
(b) A $M_0$--$H^{2/3}$ plot with the fit to Eq.~\eqref{M0}. (c) A $M_0$--$H^{1/2}$ plot with the fit to Eq.~\eqref{M00}.}
\end{figure}

The field-dependencies of $M_0$ estimated by the fit are summarized in Fig.~\ref{fig:Gap}.
According to the sine-Gordon model, $M_0$ follows the following relation\cite{HC_2},
\begin{equation}
M_0 \sim \frac{2 J v}{\sqrt{\pi}} \frac{\Gamma \Bigg(\cfrac{\xi}{2} \Bigg)}{\Gamma \Bigg( \cfrac{1+\xi}{2} \Bigg)} \left[ \frac{\Gamma \Bigg( \cfrac{1}{1+\xi} \Bigg)}{\Gamma \Bigg( \cfrac{\xi}{1+\xi} \Bigg)} \frac{c \pi h_\mathrm{s}}{2 J v} \right]^\frac{1+\xi}{2}, \label{M0}
\end{equation}
where $c$ is a parameter which is also numerically determined from Bethe ansatz.
It becomes $c \sim 1/2$ in the present case.
By using $h_\mathrm{s} = c_\mathrm{s} g_\mathrm{u} \mu_\mathrm{B} H$, Eq.~\eqref{M0} can be rewritten in a function of $H$ with only one free parameter $c_\mathrm{s}$.
The resulting function is almost proportional to ${H}^{2/3}$, because of $\xi \rightarrow 1/3$ in the vicinity of $H \rightarrow 0$\cite{1DstagDMchain, 1DstagDMchain2}.
The fit to Eq.~\eqref{M0} reproduces the field-dependencies of $M_0$ well at high fields, while they tend to underestimate $M_0$ at low fields.
Thus, the fits are performed within a fitting range of 3--9 T, which are shown as the solid curves in Fig.~\ref{fig:Gap}(a).
Good agreement can be also seen in a $M_0$--$H^{2/3}$ plot shown in Fig.~\ref{fig:Gap}(b).
A linearity of the $M_0$--$H^{2/3}$ plot supports the application of sine-Gordon model to $\mathrm{KCuMoO_4(OH)}$ above 3 T.
From the fit to Eq.~\eqref{M0}, $c_\mathrm{s}$ is estimated as 0.041(6), 0.174(6), and 0.030(6) from the fit for $H \parallel a$, $b$, and $c$, respectively.

On the other hand, for low fields, the field-dependence of $M_0$ is fitted by
\begin{equation}
M_0 \sim \sqrt{4 J S h_\mathrm{s}},
\label{M00}
\end{equation}
which is predicted by a mean field theory\cite{1DstagDMchain, 1DstagDMchain2, CMFT, CMFT2}.
The fitting curve applied to the data for $H \parallel b$ at 0--3 T is shown as a dashed curve in Fig.~\ref{fig:Gap}(a).
Although the difference between the two fits is not so large,
compared with the fit to Eq.~\eqref{M0}, that to Eq.~\eqref{M00} reproduces the field-dependencies of $M_0$ better for low fields.
This is also supported by a $M_0$--$H^{1/2}$ plot shown in Fig.~\ref{fig:Gap}(c).
Thus, the soliton mass should be proportional to $H^{1/2}$ in the vicinity of zero field while it converges into an $H^{2/3}$ dependence in high fields.
The shift of critical exponent indicates that the one-dimensional character becomes prominent in high fields
since a staggered field become so strong that the effective field produced by interchain interactions becomes negligible.
The fit to Eq.~\eqref{M00} yields $c_\mathrm{s}$ = 0.096(9) for $H \parallel b$.

In the following discussion, we rely on $c_\mathrm{s}$ derived from the fit to Eq.~\eqref{M0}
since we would like to focus on staggered fields induced by g-tensors and DM vectors inside each chain.
The $c_\mathrm{s}$ estimated from the magnetic susceptibility (fit to Eq.~\eqref {stagchi}) and the field-dependence of the soliton mass at low fields (fit to \eqref{M00})
is not appropriate for this purpose since they should be affected from staggered fields produced by interchain interactions.

Let us compare $c_\mathrm{s}$ of $\mathrm{KCuMoO_4(OH)}$ with that of other compounds listed in Table~\ref{compounds}.
In many compounds, $c_\mathrm{s}$ is about 0.1, except for 0.27 for $\mathrm{Yb_4As_3}$ due to the strong spin-orbit coupling of Yb\cite{Yb4As3_2}.
The $c_\mathrm{s}$ of $\mathrm{KCuGaF_6}$ also has a large value, 0.160 ($H \parallel b$), and 0.178 ($H \parallel c$) \cite{KCuGaF6_3}.
For $\mathrm{KCuMoO_4(OH)}$, $c_\mathrm{s}$ for $H \parallel b$ becomes a little larger than that of $\mathrm{KCuGaF_6}$,
while $c_\mathrm{s}$ for $H \parallel$ $a$ and $c$ are 4--6 times as small as that for $H \parallel b$.

\section{\label{DFT}DFT calculations}
In this section, we present results of density functional theory (DFT) calculations
that provide a deeper insight into the microscopic model of $\mathrm{KCuMoO_4(OH)}$.
In particular, an estimate of DM interactions is necessary
to understand the large variation of $c_\mathrm{s}$ depending on the field direction, as we discuss in Sec.~\ref{estimation}.
DFT calculations were performed using the generalized gradient approximation (GGA)\cite{GGA} for the exchange and
correlation potential as implemented in the full-potential code \textbf{fplo} version 14.00-47\cite{fplo}.
We used the lattice constants and atomic coordinates determined from the room temperature X-ray diffraction (XRD) measurement\cite{KCuMoO4OH, LTXRD}
except for H coordinates which have not been determined.
Thus, we placed an H atom at 1 \AA \ distance away from the O4 site and optimized its position by minimizing the GGA+$U$ total energy in the ferromagnetic arrangement.
The optimized fractional coordinates for the H site become $x/a$ = 0.229049, $y/b$ = 3/4 , and $z/c$ = 0.50401.

\begin{figure}[tb]
\includegraphics[width=8.6cm]{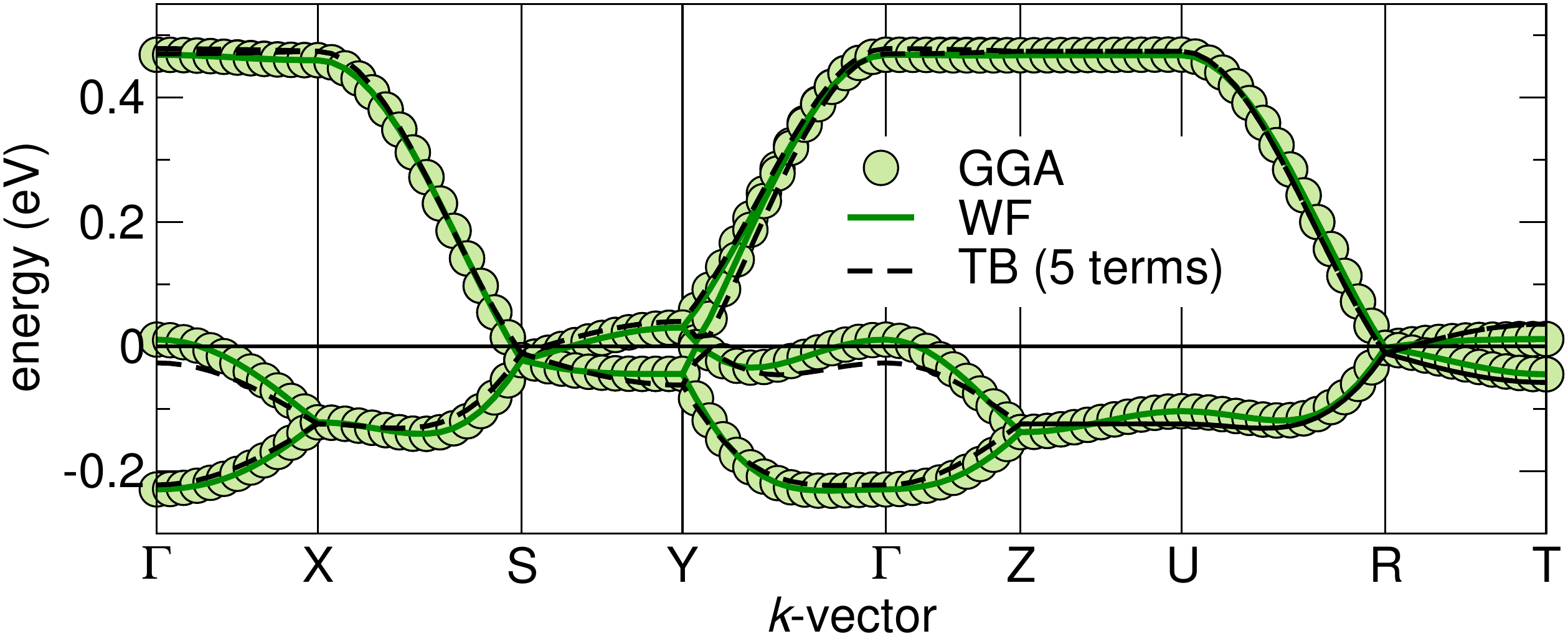}
\caption{\label{fig:bandwf} 
The GGA band structure (circles) in comparison with the Fourier-transformed
Cu-centered Wannier functions (WF) and the tight-binding model (TB)
comprising the five leading transfer integrals (Table~\ref{table:tJ}).  The
Fermi level is at zero energy. The notation of $k$-points: $\Gamma$~(000),
X~($\frac{\pi}{a}$00), S~($\frac{\pi}{a}\frac{\pi}{b}$0),
Y~(0$\frac{\pi}{b}$0), Z~(00$\frac{\pi}{c}$),
U~($\frac{\pi}{a}$0$\frac{\pi}{c}$),
R~($\frac{\pi}{a}\frac{\pi}{b}\frac{\pi}{c}$), and
T~(0$\frac{\pi}{b}\frac{\pi}{c}$).  }
\end{figure}

Then nonmagnetic band structure calculations were performed on 14 $\times$ 20 $\times $12 $k$-mesh (616 points in the irreducible wedge).
Due to the underestimation of electronic correlations, GGA yields a metallic state signaled by the presence of bands crossing the Fermi level as shown in Fig.~\ref{fig:bandwf}.
Typical for cuprates, these bands have a mixed Cu $d_{x^2-y^2}$ and O $p_\sigma$ orbital character.
This strong $dp_\sigma$ hybridization allows us to describe the low-energy physics of $\mathrm{KCuMoO_4(OH)}$ within an effective one orbital model.
To this end, we construct Cu-centered Wannier functions with the dominant $d_{x^2-y^2}$ character,
following the procedure described in Ref.~\onlinecite{Wannier}.
In this way, we obtain a 4 $\times$ 4 Hamiltonian matrix in the Wannier basis.
A Fourier transform of this Hamiltonian yields excellent agreement with the GGA bands (Fig.~\ref{fig:bandwf}).

The problem can be further simplified by restricting the Hamiltonian to the five leading terms:
the onsite energy (82~meV), the nearest-neighbor and next-nearest-neighbor couplings ($-$150 and 46~meV, respectively) along the chains, as well as two interchain couplings, listed in Table~\ref{table:tJ}.
This simplified tight-binding model is still in good agreement with the GGA band structure (Fig.~\ref{fig:bandwf}).
In second order perturbation theory, the respective antiferromagnetic exchange amounts to $4t_{ij}^2/U_\mathrm{eff}$,
where $t_{ij}$ is a transfer integral and $U_\mathrm{eff}$ is the effective onsite Coulomb repulsion acting on the Wannier orbital.
In this way, we find that the only relevant AF exchanges are nearest-neighbor and next-nearest-neighbor couplings along the chains;
the superexchange between the chains is strongly suppressed.

\begin{table}
\caption{\label{table:tJ} DFT estimates for transfer $t_{ij}$ (in meV) or
exchange $J_{ij}$ (in K) integrals 
for intrachain nearest-neighbor (NN), next-nearest-neighbor (NNN), 
interchain NN, and NNN exchanges.
The transfer integrals parametrize an
effective one-orbital model; the numerical values are evaluated using
Cu-centered Wannier functions that describe the four GGA bands around the Fermi
level.  The antiferromagnetic superexchange $J^{\text{AF}}_{ij}$ is evaluated
as $4t_{ij}^2/U_{\text{eff}}$ with $U_{\text{eff}}$\,=\,4.5\,eV.  The total
exchange $J_{ij}$ is evaluated by mapping the GGA+$U$ total energies
($U_d$\,=\,7.5, 8.0, 8.5\,eV, $J_d$\,=\,1\,eV; we use the fully localized limit
(FLL)~\cite{LDF} for the double counting correction) for different
magnetic configurations onto a classical Heisenberg model with
$|\mathbf{S}_i|$\,=\,$\frac12$. }
\begin{ruledtabular}
\begin{tabular}{rrrrrr}
\multirow{2}{*}{$\text{coupling}$}
& \multirow{2}{*}{$t_{ij}$}
& \multirow{2}{*}{$J^{\text{AF}}_{ij}$}
& \multicolumn{3}{c}{$J_{ij}$ (DFT+$U$)} \\ \cline{4-6}
& & & 7.5\,eV & 8.0\,eV & 8.5\,eV \\ \hline
onsite & 82 \\
intrachain NN\footnote{Its interatomic distance $d_{\text{Cu..Cu}}$ is $b$/2 = 3.1578 \AA.} & $-150$& 230.8& 216.7& 196.6& 181.6\\ 
intrachain NNN\footnote{$d_{\text{Cu..Cu}}$ = $b$ = 6.3156 \AA.} &     46&  22.2&$-1.3$&$-1.8$&$-4.2$\\ 
interchain NN\footnote{$d_{\text{Cu..Cu}}$ = 6.2292 \AA.} &     13&   1.7&$-0.5$&   0.2&   2.4\\ 
interchain NNN\footnote{$d_{\text{Cu..Cu}}$ = 6.9839 \AA.} &     12&   1.4&    --&    --&    --\\ 
\end{tabular}
\end{ruledtabular}
\end{table}

An apparent limitation of the one-orbital effective model is its inherent restriction to antiferromagnetic exchange, in accord with the Pauli principle.
To estimate the total magnetic exchange which also contains the ferromagnetic contributions,
we use the total energy method: GGA+$U$ total energies computed for different magnetic configurations are mapped onto a classical Heisenberg model with $|\mathbf{S}_i| =1/2$.
The respective exchange integrals are then obtained by solving a redundant system of linear equations.
Since the exact value of the onsite Coulomb repulsion $U_d$ is not known,
we performed calculations for different $U_d$ in the range 7.5--8.5 eV, by keeping the onsite Hund's exchange fixed to $J_d$ = 1 eV.
The resulting values of the exchange integrals are provided in Table~\ref{table:tJ}.
The main result of the GGA+$U$ calculations is the suppression of the next-nearest-neighbor coupling, which becomes weak and ferromagnetic.
Thus, 
$\mathrm{KCuMoO_4(OH)}$ is well regarded as a nearest-neighbor Heisenberg chain model.
Three sets of GGA+$U$ calculations performed for different $U_d$ values allow us to estimate the error bars for the leading coupling ($\sim\pm{}20$\,K).  Hence, also the magnitude of $J\sim{}200\pm{}20$\,K is in quantitative agreement with the experimental estimate.

Next, we estimate the leading anisotropies which are detected through magnetization and heat capacity measurements.
We performed non-collinear DFT+$U$ calculations implemented in the projector-augmented wave code \textbf{vasp} version 5.2.12\cite{vasp, vasp2}.
The experimental value of $J$ can be reproduced in the rotationally invariant version of DFT+$U$\cite{DFTU}
with $U_d$ = 8.5~eV and $J_d$ = 1~eV and the FLL double counting correction.
The anisotropic terms are evaluated by using the four-state mapping method for full-relativistic DFT+$U$ total energies\cite{relativisticDFT}.

As discussed in the previous section, the leading anisotropy for the nearest-neighbor exchange is the antisymmetric Dzyaloshinsky-Moriya (DM) interactions,
while further anisotropic terms are below the precision of the computational method.
Here we discuss DM interactions defined as $H_{\mathrm{DM}} = \mathbf{D}_{12} \cdot (\mathbf{S}_1 \times \mathbf{S}_2)$
between Cu atoms with fractional coordinates (1/2, 1/2, 0) (for $\mathbf{S}_1$) and (1/2, 1, 0) (for $\mathbf{S}_2$), respectively.
The discussion does not change even if different Cu sites are selected since all DM vectors are related to each other by symmetry (see Appendix~\ref{appendixA} for details). 
Due to the mirror operation with respect to the $ac$-plane and the twofold screw around the $b$-axis, 
the $b$-component of all DM vectors is 0, and the signs of $a$- and $c$- components are alternated along the chain.
The DFT+$U$ calculations yield $D_{12a}$ = 12.8 K ($D_{12a}/J = 0.054$ for $J$ = 238 K) and $D_{12c}$ = 22.0 K ($D_{12c}/J = 0.092$).
The direction of the respective DM vector is illustrated in Fig.~\ref{fig:Ddirection}(a).
For the direction of the DM vector, a guide to the eye is
the $\mathrm{MoO_4}$ tetrahedron: the DM vector points to the middle of its O--O edge spanning the neighboring
$\mathrm{CuO_6}$ octahedra in one chain. Note that in the other chain the DM vector points to the opposite direction of $\mathrm{MoO_4}$ tetrahedron because of its definition.

\section{\label{estimation}Estimation of staggered fields}
In this section, we discuss the magnitudes of $g$-tensors and DM interactions
based on $c_\mathrm{s}$ estimated by heat capacity measurements and $\mathbf{D}_0$ by DFT calculations.
We follow the notation of $\mathbf{g}_\mathrm{u}$, $\mathbf{g}_\mathrm{s}$, and  $\mathbf{D}_0$ used in \eqref{basicHamiltonian}.
From symmetry considerations, they are restricted to the following form:
\begin{equation}
\begin{split}
\mathbf{g}_\mathrm{u} &= \begin{pmatrix} g_{aa} & 0 & g_{ac} \\ 0 & g_{bb} & 0 \\ g_{ac} & 0 & g_{cc} \end{pmatrix}, \ 
\mathbf{g}_\mathrm{s} = \begin{pmatrix} 0 & g_{ab} & 0 \\ g_{ab} & 0 & g_{bc} \\ 0 & g_{bc} & 0 \end{pmatrix}, \\
\mathbf{D}_0 &= \begin{pmatrix} D_a \\ 0 \\ D_c \end{pmatrix}.
\label{definition}
\end{split}
\end{equation}
By substituting Eqs.~\eqref{definition} for $\mathbf{g}_\mathrm{u}$, and $\mathbf{g}_\mathrm{s}$ in Eqs.~\eqref{stagfield},
a uniform field $\mathbf{h}_\mathrm{u}$ and a staggered field $\mathbf{h}_\mathrm{s}$ can be described as a function of a magnetic field (see Appendix A for details).
From experimentally determined $c_\mathrm{s}$ (Fig.~\ref{fig:Gap}),
\begin{equation}
\begin{split}
\Bigg| \frac{g_{ab}}{g_{aa}} - \frac{D_c}{2J} \Bigg| &= 0.041, \\
\sqrt{\Bigg(\frac{g_{ab}}{g_{bb}} + \frac{D_c}{2J} \Bigg)^2 + \Bigg(\frac{g_{bc}}{g_{bb}} - \frac{D_a}{2J} \Bigg)^2} &= 0.174, \\
\Bigg| \frac{g_{bc}}{g_{cc}} + \frac{D_a}{2J} \Bigg| &= 0.030, \label{Habc}
\end{split}
\end{equation}
are obtained, where $J$, $g_{aa}$, $g_{bb}$, and $g_{cc}$ are given in Table~\ref{tab:param}.
Note that Eqs.~\eqref{Habc} themselves cannot be solved since they include four unknown parameters: $g_{ab}$, $g_{bc}$, $D_a/J$, and $D_c/J$, .

\begin{figure}[t]
\centering
\includegraphics[width=8cm]{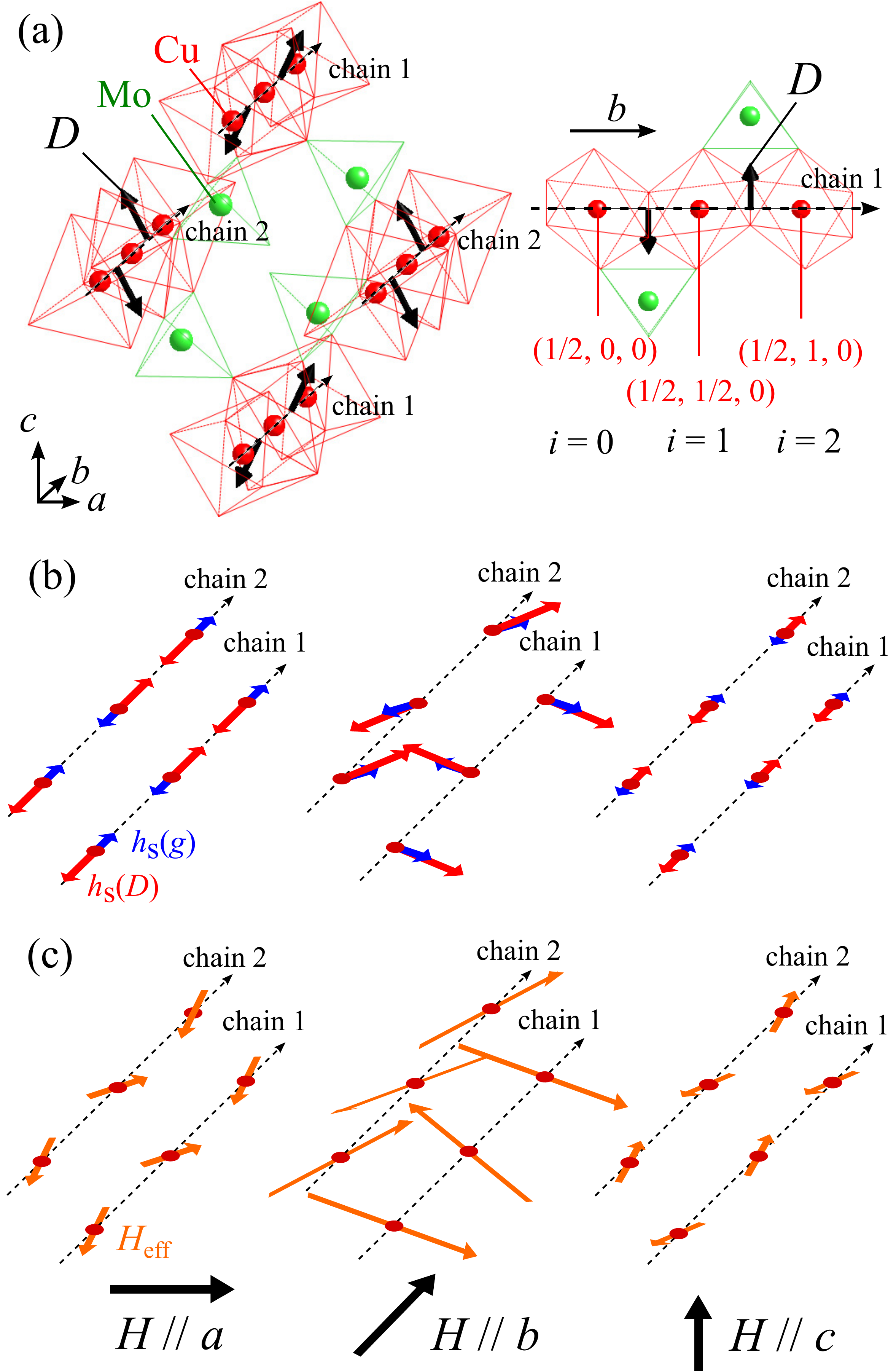}
\caption{\label{fig:Ddirection} (a) Direction of DM vectors in $\mathrm{KCuMoO_4(OH)}$.
Dashed arrows indicate the order of Cu atoms ($i$, $i+1$, $i+2$, $\cdots$) to define DM vectors.
There are four different DM vectors which are related to each other by symmetry.
(b) Schematic view of of staggered fields $\mathbf{h}_\mathrm{s}$($g$) and $\mathbf{h}_\mathrm{s}$($D$) caused from staggered $g$-tensor and DM interactions, respectively. The chains 1 and 2 are equivalent to those in Fig.~\ref{fig:Ddirection}(a).
(c) Effective magnetic fields as a sum of a uniform and staggered fields on each Cu site.}
\end{figure}

To solve Eqs.~\eqref{Habc}, the number of adjustable parameters is reduced by fixing the direction of the DM vector based on the results of DFT calculations.
For this purpose, we define $i$-th Cu atom as the Cu atom with a fractional coordinate (1/2, $i$/2, 0) (see Fig.~\ref{fig:Ddirection}(a)).
This definition leads to $\mathbf{D}_0 = \mathbf{D}_{2i-1, 2i} = -\mathbf{D}_{2i, 2i+1}$,
where $\mathbf{D}_{2i-1, 2i}$ is a DM vector defined between Cu atoms placed at $(1/2, i-1/2, 0)$ and $(1/2, i, 0)$.
Then we can introduce constraints described by
\begin{equation}
\frac{D_a}{D_c} = \frac{12.8}{22.0}.
\label{Dmratio}
\end{equation}
The parameters $g_{ab}$, $g_{bc}$, $D_a/J$, and $D_c/J$ can be determined by combining Eqs.~\eqref{Habc} and \eqref{Dmratio}.
With a help of a rough estimation on a $g$-tensor (see Appendix B for details), they are estimated as 0.129, $-$0.057, 0.110, and 0.189, respectively.
The magnitude of DM interactions is twice as large as those estimated from DFT calculations.

The magnitude of the DM vector can be also estimated from spontaneous magnetization, using the following semi-classical approach.
The spontaneous magnetization $M_\mathrm{s}$ for a two-sublattice canted-antiferromagnetic order is written as $M_\mathrm{s} = M_0 \sin \theta$, where $M_0$ and $\theta$ are the magnitude of ordered moments and their canted angle, respectively.
In addition, $\theta$ is related with a DM vector as $\tan 2\theta = |\mathbf{D}_0|/J$ according to a mean field approximation.
Thus, by using $M_\mathrm{s}$ = $0.01~\mu_\mathrm{B}$ determined from the $M$-$H$ curve and $M_0$ $\sim 0.09~\mu_\mathrm{B}$ from ordered moments of $\mathrm{Ca_2CuO_3}$\cite{Sr2CuO3_2}, which has the almost the same $T_\mathrm{N}/J$ ratio with $\mathrm{KCuMoO_4(OH)}$,
we can simply estimate the DM vector as $|\mathbf{D}_0|/J$ $\sim 0.23$, which is quite consistent with the value of $^t\mathbf{D}_0/J = (0.110, 0, 0.189)$.
Note that this semi-classical approach may lead to an underestimate since a spontaneous magnetization is measured for the unoriented sample.

From the above estimation, we can understand why the soliton mass for $H \parallel b$ becomes much larger than that for $H \parallel a$ and $c$.
To explain this, we define $\mathbf{h}_\mathrm{s}$ caused by $g$-tensor and DM interactions as $\mathbf{h}_\mathrm{s}$($g$) and $\mathbf{h}_\mathrm{s}$($D$), respectively, and illustrate them schematically in Fig.~\ref{fig:Ddirection}(b).
Although their directions vary with the Cu site, they do not have longitudinal components due to crystal symmetry.
For $H \parallel b$, four Cu sites with different combinations of $\mathbf{h}_\mathrm{s}$($g$) and $\mathbf{h}_\mathrm{s}$($D$) are present. 
In every Cu site, $\mathbf{h}_\mathrm{s}$($g$) and $\mathbf{h}_\mathrm{s}$($D$) point almost the same direction (more precisely, they are at an angle of  6$^\circ$).
This specific condition as well as large $J$ is the reason why a soliton mass becomes so large as 27 K (9 T) for $H \parallel b$.
On the other hand, for $H \parallel a$ and $c$, there are 2 Cu sites which have opposite $\mathbf{h}_\mathrm{s}$($g$) (and $\mathbf{h}_\mathrm{s}$($D$)) to each other.
In each Cu site, they point opposite to each other and the soliton mass becomes very small.
As a result, the magnitude of the effective field on each Cu site varies largely with the field direction, as shown in Fig.~\ref{fig:Ddirection}(c).
Although such a large variation of a soliton mass depending on the field direction is also observed in Cu-benzoate\cite{HC}, CDC\cite{CDC3}, and [$\mathrm{PM \cdot Cu(NO_3)_2 \cdot (H_2O)_2]_n}$] (PM = pyrimidine)\cite{PMCH},
the variation in $\mathrm{KCuMoO_4(OH)}$ is clearer owing to good crystal symmetry, large $J$, and good one-dimensionality.
Further electron spin resonance (ESR) measurements on pure or Zn-doped single crystalline samples should lead to a detailed understanding of field-induced excitations,
including boundary resonant modes observed in $\mathrm{KCuGaF_6}$ and Cu-PM\cite{BR}.

\section{\label{conclusions}Conclusions}
We investigated magnetic and thermodynamic properties of the $S$ = 1/2 quasi one-dimensional antiferromagnet $\mathrm{KCuMoO_4(OH)}$
and discussed the effect of staggered $g$-tensor and DM interactions in the presence of magnetic fields. 
The temperature-dependencies of heat capacity show good agreement with those expected in sine-Gordon model,
and from their field variations a staggered field normalized by a uniform field $c_\mathrm{s}$ is estimated as 0.041, 0.174, and 0.030 for $H \parallel a$, $b$, and $c$, respectively.
Such a large variation of $c_\mathrm{s}$ on the field direction is well understood as a combined effect of a staggered $g$-tensor and DM interactions.
They induce staggered fields in the almost same direction for $H \parallel$ $b$, leading to the large soliton mass, 
while they induce staggered fields in the opposite direction for $H \parallel$ $a$ and $c$, which make the soliton mass much smaller.

\begin{acknowledgments}
The computational results presented have been achieved in part using the Vienna Scientific Cluster (VSC).
We thank M. Oshikawa for a help to make calculations using the thermodynamic Bethe ansatz method,
and M. Oshikawa, H. Tanaka, H. Ohta, S. Okubo, and T.~J.~Sato for fruitful discussions.
The research of  K. N. was supported by Grant-in-Aid for Young Scientists (B) (No. 15K17693).
O. J. was supported by the Austrian Science Fund (FWF) through the Lise Meitner programme, project no. M2050.
\end{acknowledgments}

\appendix
\section{\label{appendixA}Calculation of a staggered field}
In this appendix, we present a calculation of a staggered field.
By substituting Eqs.~\eqref{definition} for Eqs.~\eqref{stagfield},
a uniform field $^t\mathbf{h}_\mathrm{u} \equiv (h_{ua}, h_{ub}, h_{uc})$ and a staggered field $^t\mathbf{h}_\mathrm{s} \equiv (h_{sa}, h_{sb}, h_{sc})$
are given as a function of a magnetic field $^t\mathbf{H} \equiv (H_a, H_b, H_c)$ as
\begin{equation}
\begin{split}
h_{ua} &\sim g_{aa} H_a + g_{ac} H_c, \\
h_{ub} &\sim g_{bb} H_b, \\
h_{uc} &\sim g_{ac} H_a + g_{cc} H_c, \\
h_{sa} &\sim \Bigg( g_{ab}  + g_{bb} \cfrac{D_c}{2J} \Bigg) H_b, \\
h_{sb} &\sim \Bigg( g_{ab}  - g_{aa} \cfrac{D_c}{2J} + g_{ac} \cfrac{D_a}{2J} \Bigg) H_a \\
&\ \ \ + \Bigg( g_{bc} + g_{cc} \cfrac{D_a}{2J} - g_{ac} \cfrac{D_c}{2J} \Bigg) H_c, \\
h_{sc} &\sim \Bigg( g_{bc} - g_{bb} \cfrac{D_a}{2J} \Bigg) H_b.
\label{stagfield_cal}
\end{split}
\end{equation}
Thus, $H \parallel a$ leads to $\mathbf{h}_\mathrm{u} \parallel ac$, $\mathbf{h}_\mathrm{s} \parallel b$ and 
\begin{equation}
\begin{split}
h_{ua} &\sim g_{aa} H_a, \\
h_{uc} &\sim g_{ac} H_a, \\
h_{sb} &\sim \Bigg(  g_{ab} - g_{aa} \cfrac{D_c}{2J} + g_{ac} \cfrac{D_a}{2J} \Bigg) H_a, \\
c_\mathrm{s} &\sim \frac{1}{\sqrt{g_{aa}^2 + g_{ac}^2}} \Bigg| g_{ab} - g_{aa} \frac{D_c}{2J} + g_{ac} \frac{D_a}{2J} \Bigg|. \label{Ha}
\end{split}
\end{equation}
Similarly, $H \parallel c$ leads to $\mathbf{h}_\mathrm{u} \parallel ac$, $\mathbf{h}_\mathrm{s} \parallel b$, and
\begin{equation}
\begin{split}
h_{ua} &\sim g_{ac} H_c, \\
h_{uc} &\sim g_{cc} H_c, \\
h_{sb} &\sim \Bigg( g_{bc} + g_{cc} \cfrac{D_a}{2J} - g_{ac} \cfrac{D_c}{2J} \Bigg) H_c, \\
c_\mathrm{s} &\sim \frac{1}{\sqrt{g_{ac}^2 + g_{cc}^2}} \Bigg| g_{bc}  + g_{cc} \frac{D_a}{2J} - g_{ac} \frac{D_c}{2J} \Bigg|, \label{Hc}
\end{split}
\end{equation}
and $H \parallel b$ leads to $\mathbf{h}_\mathrm{u} \parallel b$, $\mathbf{h}_\mathrm{s} \parallel ac$, and
\begin{equation}
\begin{split}
h_{ub} &\sim g_{bb} H_b, \\
h_{sa} &\sim \Bigg( g_{ab} + g_{bb} \cfrac{D_c}{2J} \Bigg) H_b, \\
h_{sc} &\sim \Bigg( g_{bc} - g_{bb} \cfrac{D_a}{2J} \Bigg) H_b, \\
c_\mathrm{s} &\sim \sqrt{\Bigg( \frac{g_{ab}}{g_{bb}} + \frac{D_c}{2J} \Big)^2 + \Big( \frac{g_{bc}}{g_{bb}} - \frac{D_a}{2J} \Bigg)^2}. \label{Hb}
\end{split}
\end{equation}
By neglecting second and higher order terms with respect to non-diagonal elements of $g$-tensor and $D/J$, $c_\mathrm{s}$ can be described in the simple form, 
\begin{equation}
\begin{split}
H \parallel a : c_\mathrm{s} &\sim \Bigg| \frac{g_{ab}}{g_{aa}} - \frac{D_c}{2J} \Bigg|, \\
H \parallel b : c_\mathrm{s} &\sim \sqrt{\Bigg( \frac{g_{ab}}{g_{bb}} + \frac{D_c}{2J} \Bigg)^2 + \Bigg(\frac{g_{bc}}{g_{bb}} - \frac{D_a}{2J} \Bigg)^2}, \\
H \parallel c : c_\mathrm{s} &\sim \Bigg| \frac{g_{bc}}{g_{cc}} + \frac{D_a}{2J} \Bigg|. \label{Habcapp}
\end{split}
\end{equation}
The above calculation applies to all four Cu sites in the unit cell of $\mathrm{KCuMoO_4(OH)}$.
This is because they are related to each other by the mirror operation with respect to the $ac$-plane and the twofold screw around the $b$-axis,
which are preserved even in the presence of the magnetic field applied to $a$, $b$, and $c$-directions.  
Note that all four Cu sites have different sets of parameters with different signs:
($g_{ab}$, $g_{bc}$, $D_a/J$, $D_c/J$), ($-g_{ab}$, $-g_{bc}$, $-D_a/J$, $-D_c/J$), ($-g_{ab}$, $g_{bc}$, $D_a/J$, $-D_c/J$), and  ($g_{ab}$, $-g_{bc}$, $-D_a/J$, $D_c/J$).
It is also obvious from Eqs.~\eqref{Habcapp} that $c_\mathrm{s}$ is unchanged whichever set is selected.

By combining Eqs.~\eqref{Habc} and Eqs.~\eqref{Dmratio}, eight solutions are obtained: ($g_{ab}$, $g_{bc}$, $D_a/J$, $D_c/J$)  
$= \pm$(0.217, $-0.134$, 0.057, 0.099), $\pm$(0.241, $-0.010$, 0.069, 0.118), $\pm$(0.091, $-0.174$, 0.092, 0.158), $\pm$(0.129, $-0.057$, 0.110, 0.189).
Among them, (0.129, $-0.057$, 0.110, 0.189) is the most likely, where $g_{ab}$ and $g_{bc}$ are the closest to $g_{ab} = 0.130$ and $g_{bc} = -0.053$ obtained from rough estimation of a $g$-tensor (see Appendix~\ref{appendixB}).\\

\section{\label{appendixB}Estimation of non-diagonal elements of $g$-tensor}
\begin{figure}[t]
\centering
\includegraphics[width=5cm]{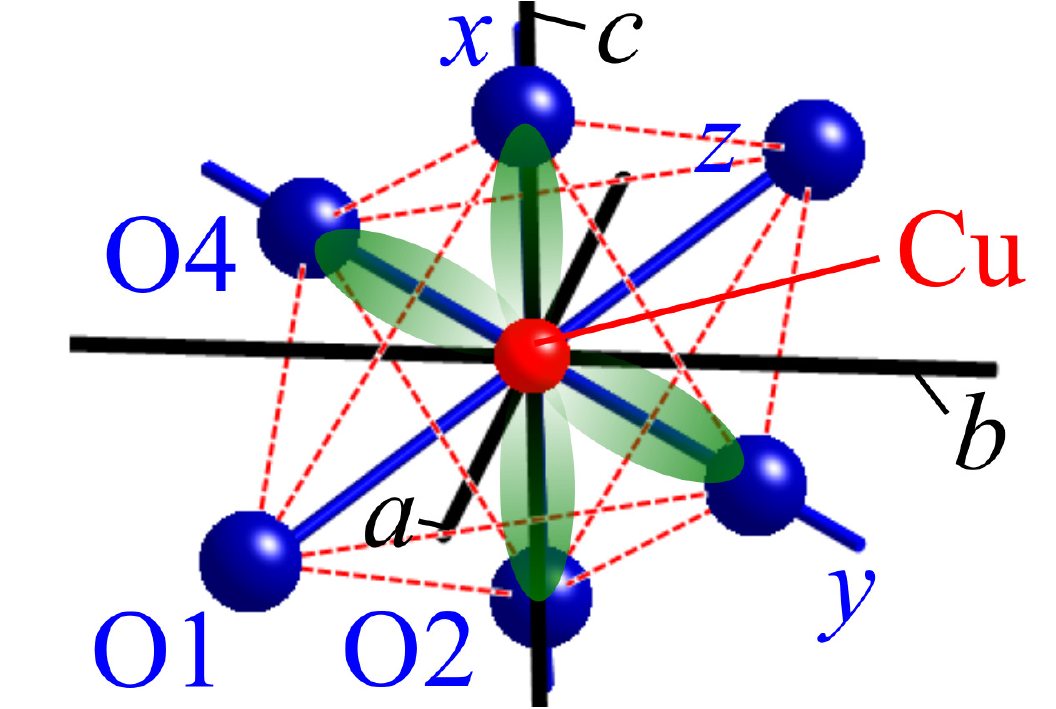}
\caption{\label{rotation} Schematic view of two different Cartesian coordinate systems $xyz$ and $abc$ defined at Cu atom (1/2, 1/2, 0).
The $x$, $y$, and $z$-axes are defined based on the local environment of the Cu site:
$x$-axis is defined along Cu--O2 bonds, $y$-axis is defined along Cu--O4 bonds, and $z$-axis is defined perpendicular to the both axes.
The $a$, $b$, and $c$-axes represent crystallographic axes of $\mathrm{KCuMoO_4(OH)}$.
Green ellipses schematically illustrate $d_{x^2-y^2}$ orbitals responsible for $S$ = 1/2.}
\end{figure}

Here we roughly estimate non-diagonal elements of $g$-tensor by multiplying rotation matrices on a typical $g$-tensor for Cu$^{2+}$.
A Cu atom is located in a distorted octahedral site surrounded by six O atoms.
Among them, O2 and O4 sites are closer to the center Cu atom than O1 sites.
Thus, we can simply regard that $d_{x^2-y^2}$ orbitals responsible for $S$ = 1/2 extend toward O2 and O4 sites as shown in green ellipses in Fig.~\ref{rotation}\cite{KCuMoO4OH}.
To describe this $g$-tensor, we need to define a Cartesian coordinate system $xyz$ where $x$ and $y$ axes are defined as the line connecting O2--Cu--O2, and O4--Cu--O4, respectively. 
The $z$-axis is defined perpendicular to the both axes, and thus O1 site is not on the $z$-axis; Cu--O1 bond forms an angle of 7.4$^\circ$ with $z$-axis.
In this coordinate system, a typical $g$-tensor for Cu$^{2+}$ should be like,
\begin{equation}
\mathbf{g} = \begin{pmatrix} 2.0 & 0 & 0 \\ 0 & 2.0 & 0 \\ 0 & 0 & 2.3 \end{pmatrix}_{xyz}
\label{gtensor}.
\end{equation}

In the following discussion, we focus on the Cu atom with fractional coordinate (1/2, 1, 0) ($i$ = 2 according to the definition in Sec.~\ref{estimation}).
Its $g$-tensor $\mathbf{g}_2$ is related with $\mathbf{g}_\mathrm{u}$ and $\mathbf{g}_\mathrm{s}$ as $\mathbf{g}_2$ = $\mathbf{g}_\mathrm{u}$ + $\mathbf{g}_\mathrm{s}$.
If we define the Cartesian coordinate system $abc$ where $a$, $b$, and $c$ axes are defined along crystallographic axes $a$, $b$, and $c$, respectively,
a rotation matrix transferring the coordinate system $xyz$ into $abc$ is given by\cite{ESR},
\begin{equation}
\mathbf{R} = \begin{pmatrix} 0.4060 & -0.5022 & -0.7635 \\ -0.0494 & 0.8222 & -0.5671 \\ 0.9126 & 0.2679 & 0.3090 \end{pmatrix}
\label{rotationmatrix}.
\end{equation}
Thus, $g$-tensor is transferred into,
\begin{equation}
\mathbf{R} \mathbf{g} \mathbf{R}^{-1} = \begin{pmatrix} 2.175 & 0.130 & -0.071 \\ 0.130 & 2.096 & -0.053 \\ -0.071 & -0.053 & 2.029 \end{pmatrix}_{abc}
\label{gtensor2},
\end{equation}
in the coordinate system $abc$.
Thus, $ab$- and $bc$-components of $\mathbf{g}_\mathrm{s}$ can be roughly estimated as $0.130$ and $-0.053$, respectively.

The same discussion applies to the $g$-tensor defined on the other Cu sites.
The estimated $g$-tensor is almost the same except that the sign of the certain non-diagonal elements is opposite.
For instance, $g$-tensor on the neighboring Cu atom (1/2 1/2, 0) ($i$ = 1) contains $ab$- and $bc$-components with the opposite sign as,
\begin{equation}
\mathbf{R} \mathbf{g} \mathbf{R}^{-1} = \begin{pmatrix} 2.175 & -0.130 & -0.071 \\ -0.130 & 2.096 & 0.053 \\ -0.071 & 0.053 & 2.029 \end{pmatrix}_{abc}
\label{gtensor3}.
\end{equation}

\begin{thebibliography}{99}
\bibitem{sign} Note that a sign in front of the DM vector $\mathbf{D}_0$ is different from the refs. \onlinecite{1DstagDMchain, 1DstagDMchain2} and modified to a corrected form in Eq.~\eqref{stagfield}.
\bibitem{LTXRD} Although the low-temperature crystal structure of $\mathrm{KCuMoO_4(OH)}$ is unknown, our thermodynamical measurements do not indicate any structural phase transitions.  Hence, DFT results for the room temperature structure should be representative for the entire temperature range studied.
\end{thebibliography}
\end{document}